\documentclass[epj,final]{svjour}
\usepackage{latexsym}
\usepackage{graphics}
\usepackage{amssymb}
\usepackage{amsmath}
\usepackage{cite}
\begin{document}
\title{An exact solution of the moving boundary problem for the relativistic plasma
expansion in a dipole magnetic field}
\author{H.B. Nersisyan\inst{1,2}\thanks{\email{hrachya@irphe.am}} \and K.A. Sargsyan\inst{1} \and D.A. Osipyan\inst{1}
\and H.H. Matevosyan\inst{1}  
}                     
%
%
\institute{Plasma Theory Group, Institute of Radiophysics and Electronics, 0203
Ashtarak, Armenia \and Centre of Strong Fields Physics, Yerevan State University, Alex Manoogian
str. 1, 0025 Yerevan, Armenia}
\date{Received: 20 April 2012 / Received in final form 16 May 2012 \\
Published online xx June 2012 -- \copyright \ EDP Sciences, Societ\`{a} Italiana di Fisica, Springer-Verlag 2012}

\abstract{
An exact analytic solution is obtained for a uniformly expanding, neutral, highly conducting plasma sphere
in an ambient dipole magnetic field with an arbitrary orientation of the dipole moment in the space. Based
on this solution the electrodynamical aspects related to the emission and transformation of energy have been
considered. In order to highlight the effect of the orientation of the dipole moment in the space we compare
our results obtained for parallel orientation with those for transversal orientation. The results obtained
can be used to treat qualitatively experimental and simulation data, and several phenomena of astrophysical
and laboratory significance.
}

\titlerunning{An exact solution for the plasma expansion in a dipole magnetic field}

\maketitle

\section{Introduction}
\label{sec:int}

There are many processes in physics which require the solution of boundary and initial value problems. When the
boundary is immobile the standard techniques for the solution of the problem are well known. However, the situation
becomes very complicated introducing a moving boundary into the problem. Usually this excludes the achievement of
an exact solution of the problem and the development of some approximate methods is desirable \cite{rog89,mor53,jac75}.
The simplest case often allowing an analytical solution is the moving plane boundary. With the help of a time-dependent
transformation it is possible to immobilize the boundary but increasing essentially the complexity of the underlying
dynamical equations. The purpose of this work is to present an example of an exactly solvable moving boundary and
initial value problem and generalize the results of our previous paper \cite{ner06}.

The problems with moving boundary are very important in fundamental mathematical physics but they have also important
applications in many area of physics. For instance, an important example is the expansion of a plasma cloud with a sharp
boundary across an ambient magnetic field. Such kind of processes are particularly of interest for many space and laboratory
investigations (see \cite{zak03,win05,osi03,ner09,col11} and references therein). This is a topic of intense interest
across a wide variety of disciplines within plasma physics, with applications to laser generated plasmas \cite{col11},
solar \cite{low01} and magnetospheric \cite{hae86,ber92} physics, astrophysics \cite{rem00}, and pellet injection for
tokamak refueling \cite{str00}.

The dynamics of the plasma cloud has been studied both numerically \cite{zak03,osi03,ner09,win05} and analytically
\cite{dit00,ner11,ner10,rai63,kat61}. Usually the plasma is treated as a highly conducting media shielded from the
penetration of the magnetic field. Thus the magnetic field is zero inside. An exact analytic solution for a uniformly
expanding plasma sphere in an external uniform magnetic field has been obtained in \cite{rai63,kat61} both for
non-relativistic \cite{rai63} and relativistic \cite{kat61} expansions. Within one-dimensional geometry a similar
problem has been considered in \cite{dit00}. In the previous papers \cite{ner06} and \cite{ner10} we obtained
the exact solutions for the uniform relativistic expansion of the plasma sphere and cylinder in the presence of a
dipole and homogeneous magnetic fields, respectively. However, in Ref.~\cite{ner06} we have considered a somewhat
simplified situation when the dipole moment is directed to the centre of the plasma sphere thus providing an azimuthal
symmetry to the problem. In the present paper we generalize the solution obtained in \cite{ner06} considering a
similar problem of the expansion of the plasma sphere in the presence of a dipole magnetic field but with an arbitrary
orientation of the dipole moment in the space.

\section{Moving boundary and initial value problem}
\label{sec:theory}

We consider the moving boundary and initial value problem of the relativistic ($v\lesssim c$, where $v$
is the radial velocity of the sphere) expansion of the plasma sphere in the vacuum in the presence of a
dipole magnetic field. Consider the dipole with the magnetic moment $\mathbf{p}$ and a neutral infinitely
conducting plasma sphere with radius $R(t)$ located at the origin of the coordinate system. The dipole is
placed in the position $\mathbf{r}_{0}$ from the centre of the sphere ($R(t)<r_{0}$). The orientation of
the dipole moment is given by the angle $\vartheta $ between the vectors $\mathbf{p}$ and $\mathbf{r}_{0}$.
In contrast to our previous paper~\cite{ner06} we consider here an arbitrary orientation of the dipole moment
in space with arbitrary $\vartheta $. The plasma sphere has expanded at $t=0$ to its present state from a
point source located at the point $\mathbf{r}=0$. It is convenient to introduce the vector potential of
the induced and the dipole magnetic fields. The magnetic field of the dipole is given by $\mathbf{H}_{0}=%
\mathbf{\nabla }\times \mathbf{A}_{0}$, where the vector potential $\mathbf{A}_{0}$ is
\begin{equation}
\mathbf{A}_{0}=\frac{\mathbf{p}\times \left( \mathbf{r}-\mathbf{r}%
_{0}\right) }{\left\vert \mathbf{r}-\mathbf{r}_{0}\right\vert ^{3}}.
\label{eq:20}
\end{equation}

As the spherical plasma cloud expands it both perturbs the external magnetic field and generates an electric
field. Within the spherical plasma region there is neither an electric field nor a magnetic field. We shall
obtain an analytic solution of the electromagnetic field configuration for an arbitrary $\vartheta$.

When an infinitely conducting plasma sphere is introduced into a background magnetic field the plasma expands
and excludes the background magnetic field to form a magnetic cavity. The magnetic energy of the dipole in the
excluded volume, i.e., in the volume of the conducting sphere is calculated as
\begin{equation}
\int_{\Omega_{R}}\frac{H_{0}^{2}(\mathbf{r}) }{8\pi } d\mathbf{r} \equiv Q(\xi )
=Q_{\parallel}(\xi )\cos^{2} \vartheta +Q_{\perp}(\xi ) \sin^{2} \vartheta ,
\label{eq:3}
\end{equation}
where $\xi =R/r_{0}<1$, $\Omega_{R}$ is the volume of the plasma sphere, $Q_{\parallel}(\xi ) =(p^{2}/2r_{0}^{3})
\xi f_{1}(\xi )$, $Q_{\perp}(\xi ) =(p^{2}/4r_{0}^{3}) \xi f_{2}(\xi )$ and the functions $f_{1}(x)$ and
$f_{2}(x)$ are determined in Appendix~A by Eqs.~(A.3) and (A.4), respectively. Here
$Q_{\parallel}(\xi )$ and $Q_{\perp}(\xi )$ are the magnetic energies escaped from the plasma volume at
parallel ($\vartheta =0$) and transversal ($\vartheta =\pi /2$) orientations of the dipole, respectively.
The energy $Q(\xi )$ increases with decreasing $\vartheta $ and reach its maximum value at $\vartheta =0$
or $\vartheta =\pi $, that is the magnetic moment $\mathbf{p}$ is parallel or antiparallel to the symmetry
axis $\mathbf{r}_{0}$.

We now turn to solve the boundary problem and calculate the induced electromagnetic fields which arise near
the surface of the plasma sphere due to the dipole magnetic field. Since the sphere is highly-conducting the
electromagnetic fields vanish inside the sphere. In addition, the normal component of the magnetic field $H_{r}$
vanishes on the surface of the sphere. To solve the boundary problem we introduce the spherical coordinate system
$r,\theta ,\varphi$ with the $z$-axis along the vector $\mathbf{r}_{0}$ and the azimuthal angle $\varphi $ is
counted from the plane ($xz$-plane) containing the vectors $\mathbf{r}_{0}$ and $\mathbf{p}$. The spherical
coordinate $\theta$ is the angle between the radius vector $\mathbf{r}$ and $\mathbf{r}_{0}$. The vector potential
(\ref{eq:20}) at $r<r_{0}$ can alternatively be represented by the sum of Legendre polynomials. Using the summation
formulas derived in Appendix~A of Ref.~\cite{ner06} one obtaines:
\begin{eqnarray}
&&A_{0r}=\frac{p}{r_{0}^{2}}\sin \vartheta \sin \varphi \sum_{\ell
=1}^{\infty }\left( \frac{r}{r_{0}}\right) ^{\ell -1}P_{\ell }^{1}\left(
\cos \theta \right) ,  \label{eq:21a} \\
&&A_{0\theta }=\frac{p}{r_{0}^{2}}\sin \vartheta \sin \varphi \sum_{\ell
=1}^{\infty }\ell \left( \frac{r}{r_{0}}\right) ^{\ell -1}P_{\ell }\left(
\cos \theta \right) ,  \label{eq:21b} \\
&&A_{0\varphi }=\frac{p}{r_{0}^{2}}\left[ \cos \vartheta \sum_{\ell
=1}^{\infty }\left( \frac{r}{r_{0}}\right) ^{\ell }P_{\ell }^{1}\left( \cos\theta \right) \right. \label{eq:21c} \\
&&\left.+\sin \vartheta \cos \varphi \sum_{\ell =0}^{\infty }\left(
\ell +1\right) \left( \frac{r}{r_{0}}\right) ^{\ell }P_{\ell }\left( \cos
\theta \right) \right] ,  \nonumber
\end{eqnarray}%
where $P_{\ell }^{\nu }(x)$ is the generalized Legendre polynomials with $\nu =0,1$ and $P_{\ell }(x)=P_{\ell }^{0}(x)$.

Since the external region of the plasma sphere is devoid of free charge density, a suitable gauge $\nabla \cdot
\mathbf{A} =0$ allows the electric and magnetic fields to be derived from the vector potential $\mathbf{A}$. Having
in mind the symmetry of the problem with respect to the $xz$-plane it is sufficient to choose the vector potential
in the form $A_{r}=A_{0r}$, and
\begin{eqnarray}
&&A_{\theta } =A_{0\theta } +\frac{p}{r_{0}^{2}}\sin \vartheta \sin \varphi \sum_{\ell
=1}^{\infty }\mathcal{A}_{\ell }(r,t) P_{\ell }^{\prime }(\cos \theta ) ,  \label{eq:22a} \\
&&A_{\varphi } =A_{0\varphi } +\frac{p}{r_{0}^{2}}\sum_{\ell =1}^{\infty }\bigg[
\cos \vartheta \mathcal{B}_{\ell }(r,t) P_{\ell }^{1}(
\cos \theta )  \label{eq:22b}  \\
&&+\sin \vartheta \cos \varphi \mathcal{C}_{\ell }(
r,t) \frac{\partial }{\partial \theta }P_{\ell }^{1}(\cos \theta ) \bigg] ,  \nonumber
\end{eqnarray}%
and the components of the electromagnetic field are given by $E_{r}=0$ and
\begin{eqnarray}
&&H_{r}=\frac{1}{r\sin \theta }\left[ \frac{\partial }{\partial \theta }%
\left( A_{\varphi }\sin \theta \right) -\frac{\partial A_{\theta }}{\partial
\varphi }\right] ,  \label{eq:23a} \\
&&H_{\theta }=\frac{1}{r\sin \theta }\frac{\partial A_{r}}{\partial \varphi }%
-\frac{1}{r}\frac{\partial \left( rA_{\varphi }\right) }{\partial r}, \label{eq:23b-1}  \\
&&H_{\varphi }=\frac{1}{r}\frac{\partial \left( rA_{\theta }\right) }{\partial
r}-\frac{1}{r}\frac{\partial A_{r}}{\partial \theta },  \label{eq:23b-2} \\
&&E_{\theta }=-\frac{1}{c}\frac{\partial A_{\theta }}{\partial t},   \quad
E_{\varphi }=-\frac{1}{c}\frac{\partial A_{\varphi }}{\partial t} . \label{eq:23c}
\end{eqnarray}%
In Eqs.~(\ref{eq:22a}) and (\ref{eq:22b}) the prime indicates the derivative with respect to the argument
and $\mathcal{A}_{\ell }(r,t)$, $\mathcal{B}_{\ell }(r,t)$ and $\mathcal{C}_{\ell }(r,t)$ are the expansion
coefficients which are determined from the boundary and initial conditions. It is straightforward to show
that the gauge $\nabla \cdot \mathbf{A} =0$ is satisfied automatically if $\mathcal{C}_{\ell }(r,t) =
\mathcal{A}_{\ell }(r,t)$. Previously in Ref.~\cite{ner06} we have considered the case of the parallel
or antiparallel orientation of the dipole when $\vartheta =0$ or $\vartheta =\pi$. In this case
Eqs.~(\ref{eq:21a})--(\ref{eq:23c}) are essentially simplified because $A_{r}=A_{0r}=0$, $A_{\theta}=A_{0\theta}=0$
and $H_{\varphi} =E_{r}=E_{\theta}=0$. Then the problem is reduced to the evaluation of the vector potential
$A_{\varphi}$ which involves only the coefficient  $\mathcal{B}_{\ell }(r,t)$.

For arbitrary $\vartheta$ the equations for the expansion coefficients $\mathcal{A}_{\ell }(r,t)$ and
$\mathcal{B}_{\ell }(r,t)$ are obtained from the Maxwell's equations
\begin{equation}
\frac{\partial ^{2}\mathcal{A}_{\ell }}{\partial r^{2}}+\frac{2}{r}\frac{%
\partial \mathcal{A}_{\ell }}{\partial r}-\frac{\ell (\ell +1)}{r^{2}}%
\mathcal{A}_{\ell }-\frac{1}{c^{2}}\frac{\partial ^{2}\mathcal{A}_{\ell }}{%
\partial t^{2}}=0 .
\label{eq:24}
\end{equation}%
Similar equation is obtained for the quantity $\mathcal{B}_{\ell }(r,t)$. This equation is to be solved in
the external region $r>R(t)$ subject to the boundary and initial conditions. Here $R(t)$ is the plasma sphere
radius at the time $t$. The initial conditions at $t=0$ are
\begin{equation}
\mathcal{A}_{\ell }(r,0)=\mathcal{B}_{\ell }(r,0)=0 ,   \quad
\frac{\partial \mathcal{A}_{\ell }(r,0)}{\partial t}=\frac{\partial \mathcal{B}_{\ell }(r,0)}{\partial t}=0 .
\label{eq:25}
\end{equation}%
The first initial condition states that the initial value of $\mathbf{A}$ is that of a dipole magnetic field,
$\mathbf{A}(\mathbf{r},0)=\mathbf{A}_{0}(\mathbf{r})$. The second initial condition states that there is no initial
electric field. Boundary conditions should be imposed at the spherical surface $r=R(t)$ and at infinity. Because
of the finite propagation velocity of the perturbed electromagnetic field the magnetic field at infinity will
remain undisturbed for all finite times. Thus, for all finite times $\mathcal{A}_{\ell}(r,t)\to 0$ and
$\mathcal{B}_{\ell }(r,t)\rightarrow 0$\ at $r\to \infty $. The boundary condition at the expanding spherical
surface is $H_{r}=0$ or, alternatively,
\begin{equation}
\mathcal{A}_{\ell }(R,t)=-\frac{1}{\ell +1}\left( \frac{R}{r_{0}}%
\right) ^{\ell } ,  \quad
\mathcal{B}_{\ell }(R,t)=-\left( \frac{R}{r_{0}}\right) ^{\ell } .
\label{eq:26}
\end{equation}

We consider the case of the uniform expansion of the plasma sphere $R(t)=vt$ with a constant expansion velocity
$v$. This special case of the uniform expansion falls within the conical flow techniques which has been applied
previously in Refs.~\cite{kat61,ner10}. From symmetry considerations one seeks a solution for the total (i.e.,
the unperturbed potential $\mathbf{A}_{0}(\mathbf{r})$ plus the induced one) vector potential of the form
$\mathcal{A}_{\ell }(r,t)=r^{\nu }\Psi _{\ell }(\zeta )$ and $\mathcal{B}_{\ell}(r,t)=r^{\nu }\Phi _{\ell }(\zeta )$
with $\zeta =r/ct$, where $c$ is the velocity of light. Here $\Psi _{\ell }(\zeta )$ and $\Phi _{\ell }(\zeta )$
are some unknown functions and $\nu >0$. Having in mind the symmetry of the unperturbed magnetic field and also
the boundary conditions~(\ref{eq:26}) it is sufficient to choose the parameter $\nu $ as $\nu =\ell $. The equation
for the vector potential $\mathbf{A}(\mathbf{r},t)$ is obtained from the Maxwell's equations which for the unknown
functions $\Psi_{\ell }(\zeta )$ and $\Phi _{\ell }(\zeta )$\ yields the same ordinary differential equation
\begin{equation}
\zeta (1-\zeta ^{2})\Psi _{\ell }^{\prime \prime }(\zeta )+2(\ell +1-\zeta
^{2})\Psi _{\ell }^{\prime }(\zeta )=0 .
\label{eq:x1}
\end{equation}%
This equation is to be solved in the external region $r>R(t)$ subject to the boundary and initial conditions. The
boundary condition at the expanding spherical surface is $H_{r}=0$ which is equivalent to the relations $\Psi
_{\ell }(\beta )=-\frac{1}{\ell +1} (\frac{1}{r_{0}})^{\ell }$ and $\Phi_{\ell }(\beta )=-(\frac{1}{r_{0}})^{\ell }$
(see Eq.~(\ref{eq:26})) with $\beta =v/c<1$. In addition, imposing that $\mathbf{A}(\mathbf{r},t)=\mathbf{A}_{0}%
(\mathbf{r})$ at $r\geqslant ct$ we obtain another boundary conditions $\Psi _{\ell }(1)=\Phi _{\ell }(1)=0$. Thus,
the solution of Eq.~(\ref{eq:x1}) subject to the initial and boundary conditions may be finally written in the form
at $vt<r<ct$
\begin{equation}
\mathcal{A}_{\ell }(r,t)=\frac{1}{\ell +1} \mathcal{B}_{\ell }(r,t) , \quad
\mathcal{B}_{\ell }(r,t)=-\left( \frac{r}{r_{0}}\right) ^{\ell }\frac{p_{\ell }(1/\zeta
)}{p_{\ell }(1/\beta )} ,
\label{eq:x2}
\end{equation}%
$\mathbf{A}(\mathbf{r},t)=\mathbf{A}_{0}(\mathbf{r})$ at $r\geqslant ct$ and $\mathbf{A}(\mathbf{r},t)=0$ at
$r\leqslant vt$. Here
\begin{equation}
p_{\ell }(z)=2^{\ell }\ell !(z^{2}-1)^{\frac{\ell +1}{2}}P_{\ell }^{-\ell
-1}(z)=\int_{1}^{z}(\tau ^{2}-1)^{\ell } d\tau
\label{eq:34a}
\end{equation}%
and $P_{\mu }^{\nu }(z)$ are the generalized Legendre functions with $z>1$, $\mu =\ell $, and $\nu =-\ell -1$.

The electromagnetic field components are determined according to Eqs.~(\ref{eq:23a})--(\ref{eq:23c}). From
Eqs.~(\ref{eq:23a})--(\ref{eq:23c}) and (\ref{eq:x2}) it is straightforward to show that the boundary conditions
on the moving surface (see, e.g., \cite{jac75,noe71} for details), $\mathbf{E}(R)=-\frac{1}{c}[\mathbf{v}\times
\mathbf{H}(R)]$ (or $E_{\theta }(R)=\beta H_{\varphi }(R)$ and $E_{\varphi }(R)=-\beta H_{\theta }(R)$), are
satisfied automatically.

It should be emphasized that all above results are valid only for $R(t)<r_{0}$ or $t<r_{0}/v$. At the time
$t=r_{0}/v$ the plasma sphere reaches to the dipole which will be completely shielded by the plasma. Therefore
at $t\geqslant r_{0}/v$ the total electromagnetic field vanishes.

Consider now briefly the non-relativistic limit of Eqs.~(\ref{eq:22a}) and (\ref{eq:22b}) with the expansion
coefficients~(\ref{eq:x2}) recalling that $\mathcal{C}_{\ell}(r,t) =\mathcal{A}_{\ell}(r,t)$. This limit can be
obtained using at $\zeta \to 0$ and $\beta \to 0$ the asymptotic expression $p_{\ell} (1/\zeta )/p_{\ell}(1/\beta )
=(\beta /\zeta)^{2\ell +1} =(R/r)^{2\ell +1}$ (see, e.g., Ref.~\cite{gra80} for the asymptotic expansion of the
generalized Legendre functions $P_{\mu}^{\nu} (z)$ with $z>1$) which yields
\begin{eqnarray}
&&A_{\theta }(\mathbf{r},t) =A_{0\theta }(\mathbf{r})-\frac{p}{r_{0}R}\sin\vartheta \sin \varphi  \label{eq:nr1} \\
&&\times\sum_{\ell =1}^{\infty }\frac{1}{\ell +1}\left( \frac{%
r_{\ast }}{r}\right) ^{\ell +1}P_{\ell }^{\prime }(\cos \theta ),  \nonumber  \\
&&A_{\varphi }(\mathbf{r},t) =A_{0\varphi }(\mathbf{r})-\frac{p}{r_{0}R}%
\sum_{\ell =1}^{\infty }\left( \frac{r_{\ast }}{r}\right) ^{\ell +1}  \label{eq:nr2}  \\
&&\times \left[\cos \vartheta P_{\ell }^{1}(\cos \theta )
+\sin \vartheta \cos \varphi \frac{1}{\ell +1}\frac{\partial }{\partial \theta }P_{\ell }^{1}(\cos \theta )\right] ,  \nonumber
\end{eqnarray}
where $r_{\ast }=R^{2}/r_{0} <R$. The remaining $\ell$-summations in these expressions can be performed using the
summation formulas derived in Ref.~\cite{ner06}. As has been shown in \cite{ner06} the resulting induced vector potential in the
non-relativistic limit represents a sum of the dipole and the quadrupole terms. These fictitious dipole and quadrupole
('image' dipole or quadrupole) are located in the $xz$-plane inside the plasma sphere at the distance $\mathbf{r}_%
{\ast }= (R^{2}/r^{2}_{0}) \mathbf{r}_{0}$ from the centre. Finally, in the lowest order with respect to the factor
$\beta$ the components of the electric field are given by $E_{\theta} =-\beta \partial A_{\theta}/\partial R$,
$E_{\varphi} =-\beta \partial A_{\varphi}/\partial R$, where $A_{\theta}$ and $A_{\varphi}$ are determined by
Eqs.~(\ref{eq:nr1}) and (\ref{eq:nr2}), respectively.

\section{Energy balance}
\label{sec:energy}

In the problem of the plasma expansion in an ambient magnetic field it is important to study the fraction of
energy emitted and lost in the form of electromagnetic pulse propagating outward of the expanding plasma
\cite{dit00,rai63,ner06,ner10,ner11}. In this section we consider the energy balance during the plasma sphere
expansion in the presence of the magnetic dipole with arbitrary orientation thus generalizing our previous
treatment~\cite{ner06} with specific orientations $\vartheta =0,\,\pi$. It is assumed that the plasma sphere
of the zero initial radius is created at $t=0$. When it starts expanding, ambient dipole magnetic field
$\mathbf{H}_{0}$ is perturbed by the electromagnetic pulse, $\mathbf{H}^{\prime } =\mathbf{H}-\mathbf{H}_{0}$,
$\mathbf{E}$, propagating outward with the speed of light. The tail of this pulse coincides with the moving
plasma boundary $r=R(t)$ while the leading edge is at the information sphere, $r=ct$. Ahead of the leading edge,
the magnetic field is not perturbed and equals $\mathbf{H}_{0}$ while the electric field is absent.

For the evaluation of the energy balance we employ the Poynting equation
\begin{equation}
\mathbf{\nabla }\cdot \mathbf{S}=-\mathbf{j}\cdot \mathbf{E}-\frac{\partial
}{\partial t}\frac{E^{2}+H^{2}}{8\pi } ,
\label{eq:35}
\end{equation}%
where $\mathbf{S}=\frac{c}{4\pi }[\mathbf{E}\times \mathbf{H}]$ is the Poynting vector and $\mathbf{j}=j_{\theta }%
\mathbf{e}_{\theta }+j_{\varphi }\mathbf{e}_{\varphi }$ (with $|\mathbf{e}_{\theta }|=|\mathbf{e}_{\varphi}|=1$)
is the surface current density. The energy components involved in Eq.~(\ref{eq:35}) have been evaluated in detail
in \cite{ner06}. We recall here some results for completeness. In order to calculate the energy emitted to infinity
we should integrate the Poynting vector over time and over the surface $S_{c}$ of the sphere with radius $r_{c}<r_{0}$
(control sphere) and the volume $\Omega _{c}$ enclosing the plasma sphere ($r_{c}>R$ or $0\leqslant t<r_{c}/v$).
Integrating over time and over the volume $\Omega _{c}$ Eq.~(\ref{eq:35}) is represented as
\begin{equation}
W_{S}(t)=W_{J}(t)+\Delta W_{\mathrm{EM}}(t) ,
\label{eq:36}
\end{equation}%
where
\begin{eqnarray}
&&W_{S}(t)=r_{c}^{2}\int_{0}^{t} d t^{\prime }\int_{0}^{\pi }\sin \theta d\theta
\int_{0}^{2\pi }S_{r} d\varphi ,   \nonumber  \\
&&W_{J}(t)=-\int_{0}^{t} d t^{\prime}\int_{\Omega _{c}}\mathbf{j}\cdot \mathbf{E} d\mathbf{r} . \label{eq:37}
\end{eqnarray}%
Here $S_{r}=\frac{c}{4\pi }(E_{\theta }H_{\varphi }-E_{\varphi }H_{\theta })$ is the radial component of the Poynting
vector. Also $W_{\mathrm{EM}}(t)$ and $\Delta W_{\mathrm{EM}}(t)=W_{\mathrm{EM}}(0)-W_{\mathrm{EM}}(t)$ are the total
electromagnetic energy and its change (with minus sign) in a volume $\Omega _{c}$, respectively. $W_{J}(t)$ is the
energy transferred from plasma sphere to electromagnetic field and is the mechanical work with minus sign performed by
the plasma on the external electromagnetic pressure. At $t=0$ the electromagnetic fields are given by $\mathbf{H}%
(\mathbf{r},0)=\mathbf{H}_{0}(\mathbf{r})$ and $\mathbf{E}(\mathbf{r},0)=0$. Hence $W_{\mathrm{EM}}(0) $ is the energy
of the dipole magnetic field in a volume $\Omega _{c}$ and can be calculated from Eq.~(\ref{eq:3}) by replacing $R$ by
$r_{c}$. Thus $W_{\mathrm{EM}}(0) =Q(u)$, where $u=r_{c}/ r_{0}<1$. Taking into account that $\mathbf{H}=\mathbf{E}=0$
in a plasma sphere, the change of the electromagnetic energy $\Delta W_{\mathrm{EM}}(t)$ in a volume $\Omega _{c}$ can
be represented in the form \cite{ner06}
\begin{equation}
\Delta W_{\mathrm{EM}}(t)
=Q(u)-\int_{\Omega _{c}^{\prime }}\frac{E^{2}+H^{2}}{8\pi } d\mathbf{r} ,
\label{eq:39}
\end{equation}%
where $\Omega _{c}^{\prime }$ is the volume of the control sphere excluding the volume of the plasma sphere. Hence the
total energy flux, $W_{S}(t)$ given by Eq.~(\ref{eq:37}) is calculated as a sum of the energy loss by plasma due to the
external electromagnetic pressure and the decrease of the electromagnetic energy in a control volume $\Omega _{c}$.

Consider now explicitly each energy component $W_{S}(t)$, $W_{J}(t)$ and $\Delta W_{\mathrm{EM}}(t)$. $W_{S}(t)$ is evaluated
from Eq.~(\ref{eq:37}). In the first expression of Eq.~(\ref{eq:37}) the $t^{\prime }$-integral must be performed at
$\frac{r_{c}}{c}\leqslant t^{\prime}\leqslant t$ ($t<\frac{r_{c}}{v}$) since at $0\leqslant t^{\prime }<\frac{r_{c}}{c}$
the electromagnetic pulse does not reach to the control surface yet and $S_{r}(r_{c}) =0$. Using the summation formulas
derived in Appendix from Eqs.~(\ref{eq:23a})--(\ref{eq:23c}), (\ref{eq:x2}) and (\ref{eq:37}) we obtain
\begin{eqnarray}
&&W_{S}(t)=Q(u)+\frac{p^{2}}{2r_{0}^{3}}\sum_{\ell =1}^{\infty }\frac{\ell
a_{\ell }(\vartheta )}{2\ell +1}u^{2\ell +1}   \nonumber  \\
&&\times\left\{ \frac{(1/\eta ^{2}-1)^{2\ell +1}}{(2\ell +1)p_{\ell }^{2}(1/\beta )}-(\ell +1)\left[ \frac{%
p_{\ell }(1/\eta )}{p_{\ell }(1/\beta )}-1\right] ^{2}\right\} ,  \label{eq:40}
\end{eqnarray}%
where $\eta =r_{c}/ct<1$, and
\begin{equation}
a_{\ell }(\vartheta )=(\ell +1)\cos ^{2}\vartheta + \frac{\ell }{2}\sin ^{2}\vartheta
\label{eq:new}
\end{equation}
is the orientation factor of the dipole magnetic field. In the non-relativistic limit, $\beta \to 0$, using the
asymptotic expression (see, e.g., Ref.~\cite{gra80}) $p_{\ell }(z)=z^{2\ell +1}/(2\ell +1)$ at $z\to \infty $, as well
as the summation formulas of Appendix from Eq.~(\ref{eq:40}) we obtain
\begin{eqnarray}
&&W_{S}(t)=2Q(\xi )-Q(\kappa )  \nonumber  \\
&&+\frac{p^{2}}{r_{0}^{3}}\frac{\kappa ^{3}}{(1-\kappa ^{2})^{3}}\left( \cos ^{2}\vartheta
+\frac{1+\kappa ^{2}}{4}\sin^{2}\vartheta \right)  \label{eq:42}
\end{eqnarray}%
with $\kappa =R^{2}/r_{0}r_{c}$. In Eq.~(\ref{eq:42}) $Q(\kappa )$ represents the magnetic energy of the dipole
field in a sphere having the radius $R_{\ast }=R^{2}/r_{c}<R$ and enclosed in the plasma sphere.

\begin{figure*}
\begin{center}
\resizebox{1.70\columnwidth}{!}{%
  \includegraphics{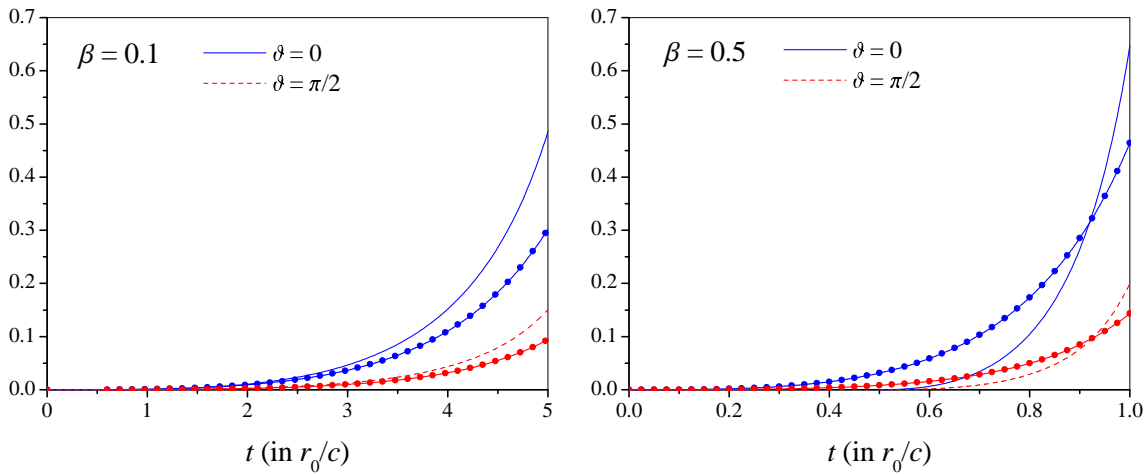}}
\end{center}
\caption{(Color online) The energy components $W_{S}(t)$ (the lines without symbols) and $W_{J}(t)$ (the lines with symbols)
normalized to the quantity $p^{2}/ r^{3}_{0}$ for two values of the factor $\beta $ as a function of $ct/r_{0}$
calculated from expressions (\ref{eq:40}) and (\ref{eq:44}) with $r_{c}=0.5r_{0}$. The solid and dashed lines
correspond to the dipole orientations $\vartheta =0$ and $\vartheta =\pi /2$, respectively.}
\label{fig:1}
\end{figure*}

Next, we evaluate the energy loss $W_{J}(t)$ by the plasma which is determined by the surface current density,
$\mathbf{j}$ localized within thin skin layer, $R-\delta <r<R+\delta $ with $\delta \to 0$, near plasma boundary.
Therefore in Eq.~(\ref{eq:37}) the volume $\Omega _{c}$ is in fact replaced by the volume $\Omega _{\delta }\sim R^{2}%
\delta $ of the shell with $R-\delta <r< R+\delta $. Within the skin layer we take into account that $\mathbf{E}=%
-\frac{1}{c}[\mathbf{v}\times \mathbf{H}]$ and $H_{r}(R)=0$. Also it should be mentioned that the boundary of the volume
$\Omega_{\delta }$ moves with a velocity $v$ and the electrical field has a jump across the plasma surface. The detailed
calculations are given in \cite{ner06} and we provide here only the final result which reads
\begin{equation}
Q_{J}(t)=-\int_{\Omega _{\delta }}\mathbf{j}\cdot \mathbf{E} d\mathbf{r}
=\frac{v}{\gamma ^{2}}\int\nolimits_{S_{R}}\frac{H^{2}(R)}{8\pi } dS .
\label{eq:43}
\end{equation}%
Here $\gamma ^{-2}=1-\beta ^{2}$ and $S_{R}$ are the relativistic factor and the surface of the expanding plasma,
respectively. Note that the moving boundary modifies the surface current which is now proportional to the factor
$\gamma^{-2}$ \cite{jac75,noe71}. Equation~(\ref{eq:43}) shows that the energy loss by the plasma per unit time is
equal to the work performed by the plasma on the external electromagnetic pressure. As shown in Ref.~\cite{ner06}
this external pressure is formed by the difference between magnetic and electric pressures, i.e., the induced
electric field tends to decrease the force acting on the expanding plasma surface. The total energy loss by the
plasma sphere is calculated as
\begin{eqnarray}
W_{J}(t) =\int_{0}^{t}Q_{J}\left( t^{\prime }\right) d t^{\prime} =\frac{p^{2}}{2r_{0}^{3}}\sum_{\ell =1}^{\infty }
\frac{\ell a_{\ell}(\vartheta )}{(2\ell +1)^{2}}  \nonumber \\
\times \left( \frac{\xi }{\beta ^{2}\gamma ^{2}}\right) ^{2\ell +1}\frac{1}{p_{\ell }^{2}(1/\beta )} , \label{eq:44}
\end{eqnarray}%
where $\xi =R/r_{0}$. In a non-relativistic case Eq.~(\ref{eq:44}) yields (see Appendix):
\begin{equation}
W_{J}(t)=\frac{p^{2}}{r_{0}^{3}}\frac{\xi ^{3}}{\left( 1-\xi ^{2}\right) ^{3}%
}\left( \cos ^{2}\vartheta +\frac{1+\xi ^{2}}{4}\sin ^{2}\vartheta \right) .
\label{eq:45}
\end{equation}

The change of the electromagnetic energy in a control sphere is calculated from Eq.~(\ref{eq:39}). At $R<r_{c}<ct$
(the electromagnetic pulse fills the whole control sphere) it is straightforward to show that $\Delta W_{\mathrm{EM}}(t)
=W_{S}(t) -W_{J}(t)$ as predicted by the energy balance equation~(\ref{eq:36}). The non-relativistic limit of
$\Delta W_{\mathrm{EM}}(t)$ can be evaluated from Eqs.~(\ref{eq:42}) and (\ref{eq:45}). Also it should be emphasized
that in Eqs.~(\ref{eq:40}), (\ref{eq:44}) as well as in $\Delta W_{\mathrm{EM}}(t)$ the dependence on the orientation
$\vartheta$ of the dipole moment is determined by the factor $a_{\ell} (\vartheta )$ given by (\ref{eq:new}). At parallel
(or antiparallel) orientation with $\vartheta =0$ (or $\vartheta =\pi$) this factor is $a_{\ell} =\ell +1$ and
Eqs.~(\ref{eq:40}), (\ref{eq:44}) as well as $\Delta W_{\mathrm{EM}}(t)$ coincide with the corresponding formulas derived
in Ref.~\cite{ner06}. As in Eq.~(\ref{eq:3}) it is convenient to represent the energy components $W_{S}(t)$ and $W_{J}(t)$
in the form $W_{S,J}(t) =W_{S,J\parallel}(t) \cos^{2} \vartheta +W_{S,J\perp}(t) \sin^{2} \vartheta$, where $W_{\parallel}(t)$
and $W_{\perp}(t)$ are the energy components at parallel ($\vartheta =0$) and transversal ($\vartheta =\pi /2$) orientations
of the dipole moment, respectively.

As an example in Figure~\ref{fig:1} we show the results of model
calculations for the time evolution of the energy components $W_{S}(t)$ and $W_{J}(t)$ for two extreme orientations
of the dipole magnetic field with $\vartheta =0$ and $\vartheta =\pi /2$. We recall that at $0\leqslant t\leqslant
r_{c}/c$, i.e. the electromagnetic pulse does not yet reach to the surface of the control sphere, the energy flux
vanishes and $W_{S}(t)=0$. Thus, for a given radius $r_{c}$ of the control sphere the energy flux occurs within
the time interval $\Delta t= (r_{c}/c) (1/\beta -1)$. Note that this interval decreases with the expansion velocity.
As expected in Figure~\ref{fig:1} it is seen that both components of the energy $W_{S}(t)$ and $W_{J}(t)$ increase
monotonically with time and strongly depend on the orientation $\vartheta$ of the magnetic dipole. In addition,
since the dipole magnetic field at the location of the plasma cloud is stronger at smaller orientation angles $\vartheta$
the energy flux $W_{S}(t)$ and the energy loss $W_{J}(t)$ are maximal for the parallel (or antiparallel) orientation
of $\mathbf{p}$. It is also noteworthy the dependence of the energy components on the relativistic factor $\beta$ of
the plasma expansion. In the case of the non-relativistic expansion the energy flux $W_{S}(t)$ is systematically larger
than the energy loss $W_{J}(t)$ (see Figure~\ref{fig:1}, left panel) at least in the wide interval of the expansion time
(for comparison see Figure~\ref{fig:2} below). Furthermore, both energies increase with the factor $\beta$ but now
$W_{J}(t)$ exceeds $W_{S}(t)$ in the wide interval of the expansion time (Figure~\ref{fig:1}, right panel).

\begin{figure*}
\begin{center}
\resizebox{1.70\columnwidth}{!}{%
  \includegraphics{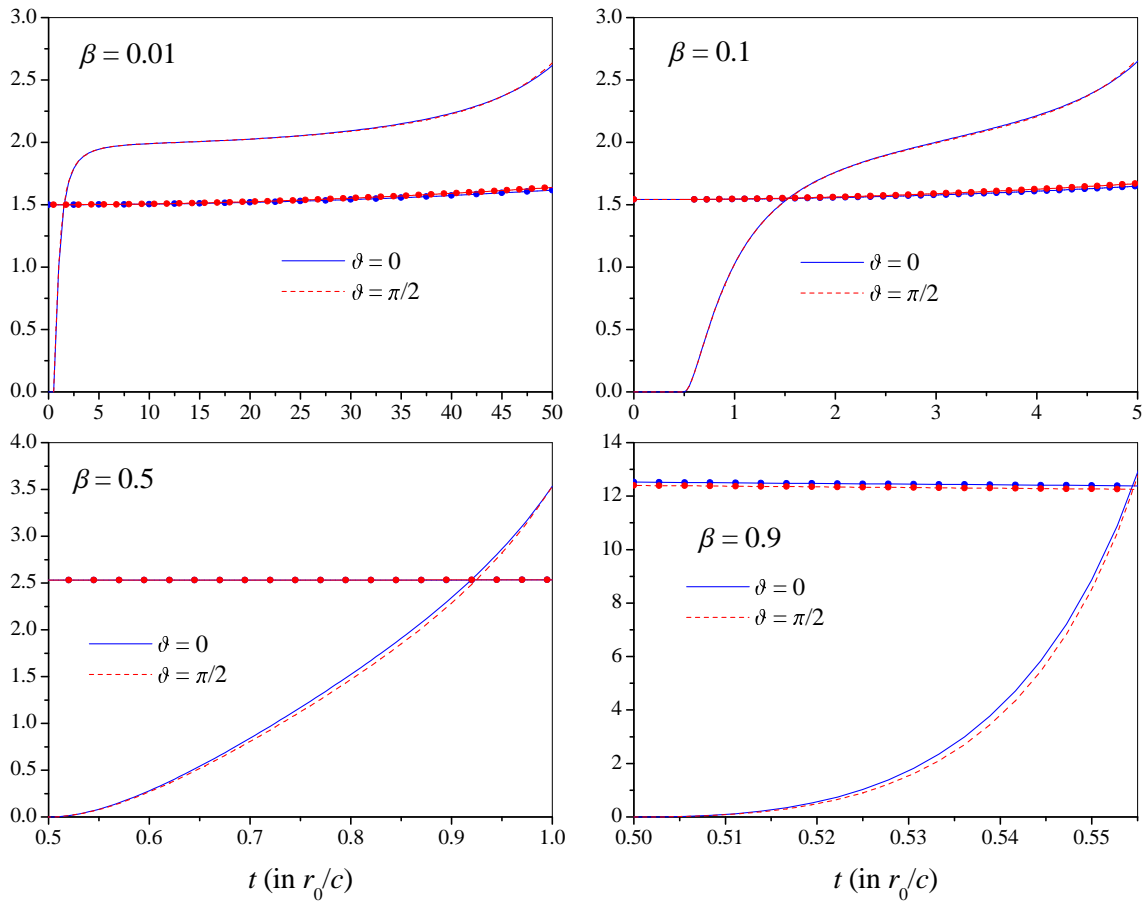}}
\end{center}
\caption{(Color online) The ratios $\Gamma _{S}(t)$ (the lines without symbols) and $\Gamma _{J}(t)$ (the lines with symbols)
for four values of the factor $\beta $ as a function of $ct/r_{0}$ calculated from expressions (\ref{eq:40}) and
(\ref{eq:44}) with $r_{c}=0.5r_{0}$. The solid and dashed lines correspond to the dipole orientations $\vartheta =0$
and $\vartheta =\pi /2$, respectively.}
\label{fig:2}
\end{figure*}

To gain more insight in Figure~\ref{fig:2} we show also the time evolution of the ratios $\Gamma _{S}(t)=W_{S}(t)/Q_{0}(t)$
and $\Gamma _{J}(t)=W_{J}(t)/Q_{0}(t)$ for $\vartheta =0$ and $\vartheta =\pi /2$. The relativistic factor $\beta $
varies from the non-relativistic ($\beta =0.01$) to the relativistic ($\beta =0.9$) values. Here $Q_{0}(t)=Q(\xi )$
is the dipole magnetic energy escaped from the plasma sphere. Previously it has been found \cite{dit00} that for the
non-relativistic ($\beta \ll 1$) expansion of a one-dimensional plasma slab and for uniform external magnetic field
approximately the half of the outgoing energy is gained from the plasma, while the other half is gained from the magnetic
energy, $W_{S}\simeq 2W_{J}\simeq 2\Delta W_{\mathrm{EM}}$. In the case of the expansion of a spherical plasma with
$\beta \ll 1$ in the uniform magnetic field, $W_{J}=1.5Q_{0}$, $\Delta W_{\mathrm{EM}}=0.5Q_{0}$, and $W_{S}=2Q_{0}$,
where $Q_{0}=H_{0}^{2}R^{3}/6$ \cite{rai63}. Therefore in this case the released electromagnetic energy is mainly gained
from the plasma.

In contrast to the cases with uniform magnetic field discussed above in the present context there are no simple relations
between the energy components $W_{S}(t)$, $W_{J}(t)$ and $Q_{0}(t)$. However, at the initial stage ($t\ll r_{c}/v$) of
non-relativistic expansion the dipole field at large distances can be treated as uniform and the energies $W_{S}(t) $ and
$W_{J}(t)$ are close to the values $2Q_{0}(t)$ and $1.5Q_{0}(t)$ (Figure~\ref{fig:2}), respectively. It is noteworthy that
for arbitrary $\beta$ the ratios $\Gamma _{S}(t)$ and $\Gamma _{J}(t)$ are only weakly sensitive to the orientation of
the dipole magnetic field although the quantities $W_{S}(t)$, $W_{J}(t)$ and $Q_{0}(t)$ strongly depend on $\vartheta$
(see Figure~\ref{fig:1} and Eq.~\eqref{eq:3} with (A.3), (A.4)). Moreover, for any $\beta $ the ratio $\Gamma _{J}(t)$ is
almost constant and may be approximated as $\Gamma _{J}(t)\simeq \Gamma _{J}(0)$ or alternatively $W_{J}(t)\simeq 1.5CQ_{0}(t)$,
where $C=\gamma ^{-6}(1-\beta )^{-4}(1+2\beta )^{-2}$ is some kinematic factor which is independent on the dipole orientation
$\vartheta$. These features merely indicate that for arbitrary $\beta$ the energy loss $W_{J}(t)$ increases proportionally
to the escaped energy $Q_{0}(t)$ while the parallel ($W_{S\parallel}(t)$) and the transversal ($W_{S\perp}(t)$) components
of the emitted energy increase proportionally to the parallel ($Q_{0\parallel}(t)$) and transversal ($Q_{0\perp}(t)$)
components of the escaped energy, respectively, with $W_{S\parallel}(t)/Q_{0\parallel}(t) \simeq W_{S\perp}(t)/Q_{0\perp}(t)%
\simeq \Gamma_{S}(t)$. Finally, in Figures~\ref{fig:1} and \ref{fig:2} it is seen that for non-relativistic expansion the
emitted energy is gained from both the plasma cloud and electromagnetic energy in the control sphere. However, at the final
stage ($t=r_{c}/v$) of the relativistic expansion (with $\beta \sim 1$) $W_{S}\simeq W_{J}$ and hence in this case the emitted
energy $W_{S}$ is mainly gained from the plasma.

\section{Conclusion}
\label{sec:conc}

An exact solution of the uniform radial expansion of a neutral, infinitely conducting plasma sphere in the
presence of a dipole magnetic field has been obtained for an arbitrary orientation of the dipole moment in
the space. Thus we have generalized our result obtained in Ref.~\cite{ner06} for a parallel (or antiparallel)
orientation of the dipole moment when $\mathbf{p} \parallel \mathbf{r}_{0}$. The electromagnetic fields are
derived by using the appropriate initial and boundary conditions, Eqs.~(\ref{eq:25}) and (\ref{eq:26}),
respectively. As expected the electromagnetic fields are perturbed only within the domain located between the
surfaces of the expanding plasma ($r=R$) and the information ($r=ct$) spheres. Outside the sphere $r=ct$ the
magnetic field is not perturbed and coincides with the dipole magnetic field. In addition, since the dipole
magnetic field at the location of the plasma cloud is stronger for parallel (or antiparallel) orientation the
induced electromagnetic fields are maximal for $\mathbf{p} \parallel \mathbf{r}_{0}$. We have also studied
the energy balance during the plasma sphere expansion with an arbitrary orientation of the dipole. The model
calculations demonstrate that the energy components increase monotonically with time and strongly depend on
the orientation of the dipole moment (Figure~\ref{fig:1}), being maximal at $\vartheta =0,\pi$. On the other
hand, the ratios $\Gamma_{S}(t)$ and $\Gamma_{J}(t)$ are only weakly sensitive to the variation of $\vartheta$,
the latter being a slowly varying function of time (Figure~\ref{fig:2}). This is because the energy loss
$W_{J}(t)$ varies almost linearly with the escaped energy $Q_{0}(t)$ while the parallel ($W_{S\parallel}(t)$)
and the transversal ($W_{S\perp} (t)$) components of the emitted energy vary linearly with the parallel
($Q_{0\parallel}(t)$) and transversal ($Q_{0\perp}(t)$) components of the escaped energy, respectively.
The calculations also show that for non-relativistic expansion the emitted energy is gained from both the plasma cloud
and perturbations of the electromagnetic energy in the control sphere. For relativistic expansion $W_{S}\simeq
W_{J}$ and the emitted energy is practically gained only from the plasma sphere.

Lastly, to solve the boundary value problem we have applied the conical flow technique which is only valid
for the uniform expansion of the plasma cloud with constant $v$. In principle the solution for the arbitrary
time-dependent velocity $v(t)$ can be obtained by the Laplace transformation technique used, for instance, in
\cite{ner06}. This yields an integral equation for the vector potential which allows a general analytical solution
only for a one--dimensional expansion. For two-- and three--dimensional expansions the general solutions are not
known and these cases require separate investigations.

\begin{acknowledgement}
This work has been supported by the State Committee of Science of Armenian
Ministry of Higher Education and Science (Project No. 11--1c317).
\end{acknowledgement}

\section*{Appendix: Some summation formulas}
\label{sec:ap1}

Using the known relation \cite{gra80}
\begin{equation}
\chi (x) =\sum_{\ell =1}^{\infty }x^{2\ell }=\frac{x^{2}}{1-x^{2}} , \tag{A.1}
\end{equation}%
where $x<1$, one can derive some summation formulas which are used in the main text of the paper. The first
relation is evaluated as
\begin{gather}
f(x) =\sum_{\ell =1}^{\infty }\frac{\ell (\ell +1) }{2\ell +1}x^{2\ell +1}
= \frac{1}{4} \left\{x\frac{\partial}{\partial x} [x\chi (x)] -\int_{0}^{x}\chi (t) dt \right\}  \nonumber \\
=\frac{1}{8}\left[ \frac{x(1+x^{2})}{(1-x^{2})^{2}}- \frac{1}{2}\ln \frac{1+x}{1-x}\right] .    \tag{A.2}
\end{gather}%
In (A.2) we have used the boundary condition $f(0)=0$.

Similarly one obtains:
\begin{gather}
f_{1}(x) =\sum_{\ell =1}^{\infty }\frac{\ell (\ell +1) ^{2}}{2\ell +1}x^{2\ell }
=\frac{1}{2x}\frac{\partial }{\partial x}\left[ xf(x) \right]  \nonumber  \\
=\frac{1}{8}\left[ \frac{8x^{2}-x^{4}+1}{(1-x^{2})^{3}}-\frac{1}{2x}\ln \frac{1+x}{1-x}\right] ,  \tag{A.3}  \\
f_{2}(x) =\sum_{\ell =1}^{\infty }\frac{\ell ^{2}(\ell +1)}{2\ell +1}x^{2\ell }
=\frac{x}{2}\frac{\partial }{\partial x}\left[ \frac{1}{x}f(x) \right]  \nonumber \\
=\frac{1}{8}\left[ \frac{x^{4}+8x^{2}-1}{(1-x^{2}) ^{3}}+\frac{1}{2x}\ln \frac{1+x}{1-x}\right] . \tag{A.4}
\end{gather}

\end{document}